\documentclass[preprint,showpacs,preprintnumbers,amsmath,amssymb]{revtex4}

\usepackage{graphicx}
\usepackage{dcolumn}

\begin{document}

\preprint{APS/123-QED}

\title{Pattern formation on the surface of cationic-anionic cylindrical aggregates}%

\author{Y. S. Velichko}
\author{M. Olvera de la Cruz}%
\email{m-olvera@northwestern.edu}%
\affiliation{Department of Material Science and Engineering,
Northwestern University, IL 60208, USA.}%

\date{\today}

\begin{abstract}
Charged pattern formation on the surfaces of self--assembled cylindrical
micelles formed from oppositely charged heterogeneous molecules such as
cationic and anionic peptide amphiphiles is investigated. The net
incompatibility~$\chi$ among different components results in the formation of
segregated domains, whose growth is inhibited by electrostatics. The
transition to striped phases proceeds through an intermediate structure
governed by fluctuations, followed by states with various lamellar
orientations, which depend on cylinder radius~$R_c$ and ~$\chi$. We analyze
the specific heat, susceptibility~$S(q^*)$, domain size~$\Lambda=2\pi/q^*$
and morphology as a function of~$R_c$ and~$\chi$.
\end{abstract}

\pacs{87.15.-v, 82.35.Pg, 81.16.Fg, 81.07.-b, 61.46.+w, 64.60.-i}
\keywords{Self--assemble, pattern formation, charges, cylindrical monolayer}
\maketitle

Many heterogenous molecules, including lipids and amphiphiles, assemble into
finite size aggregates or micelles
~\cite{ZembNature2001,KalerJCP1992,StuppJACS2003,StuppScience2004,KellerPRL2002}.
Co-assembly of cationic and anionic heterogeneous molecules is a route to
create functional biomolecular materials, such as stable vesicles for drug
delivery~\cite{ZembNature2001,KalerJCP1992} and bio--active
fibers~\cite{StuppJACS2003,StuppScience2004}. Co-assembled cationic and
anionic systems are ubiquitous in nature since most biomolecules are
heterogenous, charged and strongly associating in aqueous solutions. The
competition of long--range electrostatic and short--range interactions among
different co-assembled components may lead to the formation of complex
aggregates with surface charge heterogeneities. These heterogeneities are
expected to play a prominent role in the fabrication of functional assemblies
and in the self-organization of the aggregates.

The assembly of single component charged amphiphilic molecules is restricted
by electrostatic repulsion~\cite{HuangMacromol1997,ZhulinaMacromol2003}. For
example, electrostatics restricts assembly of charged peptide amphiphiles
(PA) composed of a hydrophobic block connected to a peptide block that favors
$\beta$-sheet formation. Single component PAs assemble only when the peptide
blocks are neutral (acidic PAs assemble at high pH and basic PAs at low pH
conditions) and/or when the salt concentration is excessively high. However,
at physiological pH conditions, when the molecules are charged,
stoichiometric mixtures of acidic $(-)$ and basic $(+)$ PAs co-assemble at
one percent concentration into nanofibers of diameter about $6-8$~nm and a
few microns length, which form a network that resembles extracellular matrix
found in living tissue~\cite{StuppJACS2003}. The peptide amphiphile charged
end groups are exposed to the surface of the fibers, and their amino acids
sequences can be designed to promote the growth of
bone~\cite{StuppScience2001}, or neural cell
differentiation~\cite{StuppScience2004}.


Co-assembled neutral molecules, such as phospholipids and cholesterol in
mono-- or bi--layers, are generally unstable. These neutral assemblies
generate domains of different density or composition along the surface, which
coarsen with time~\cite{KellerPRL2002} and, eventually, lead to the
dissolution of the co-assembled structure~\cite{LaradjiPRL2004}. Instead, in
co-assembled chemically dissimilar cationic and anionic amphiphile mixtures,
the compositional heterogeneities driven by a net incompatibility among the
different components are stabilized by electrostatics, which favors mixing of
the charged components. At low temperature $T$, the competition of the
interfacial energy (which favors large domains) and the electrostatic
interactions (which favors small domains) is expected to result in periodic
surface charged domains~\cite{SolisJCP2005} of size $\Lambda$. The cohesive
energy $|E|$ of the domains stabilizes the co-assembled periodic structure at
low $T$. In analogy with ionic crystals of ions of valence $Z$ and
periodicity $\Lambda$ at $T=0$ where $E\sim-Z^2/\Lambda$~\cite{Madelung},
since in our system $Z\sim \Lambda^2$, the energy per particle $N\sim
\Lambda^2$ scales as $E/N \sim -\Lambda$ leading to robust periodic
structures. In this letter we study the formation of charged domains in
cylindrical aggregates with Coulomb interactions at various degrees of
incompatibility and cylinder radii. We analyze charge correlations and domain
orientations, as well as the thermodynamics of the local segregation process.
We show that phase segregation on the surface of co-assembled
cationic-anionic micelles can be readily observed when the net
incompatibility among dissimilar molecules generated by chemical and/or
molecular sizes differences is larger than a couple $k_BT$'s.

\begin{figure}
\includegraphics[width=8.4cm]{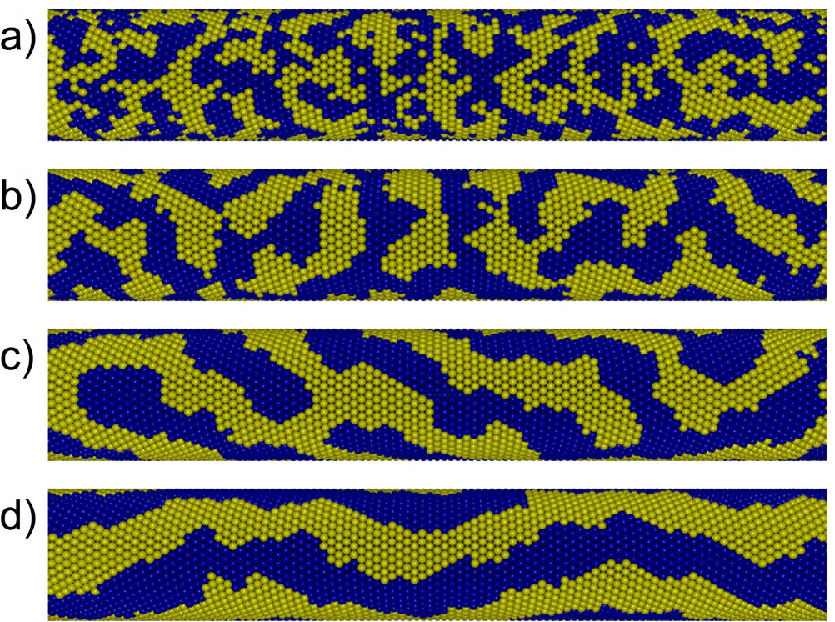}
\caption{\label{fig-1}Snapshots of typical configurations of (a)~isotropic
phase~($\chi=4.5$), (b)~locally correlated state~($\chi=6.75$), (c)~striped
state with defects~($\chi=9.0$) and (d)~parallel striped phase~($\chi=12.0$).
$R_c/\sigma=8$.}
\end{figure}

We consider each aggregate as a stable structure, thus constraining the
cylindrical geometry of the micelles and examining only equilibrium surface
structures. A binary electroneutral charged lattice fluid consisting of
positive and negative units of equal absolute charge $|Q_+|=|Q_-|$, confined
to a cylindrical monolayer of radius $R_c$ and length $L_c=L_z$, is placed in
the center of the box $L\times L\times L_z$ along the $z-$axis. The surface
of the cylinder is filled with spherical units of diameter $\sigma=1,$ in
such a way, that all units are placed into the knots of a triangular lattice
of period~$a=\sigma$~(Fig.~\ref{fig-1}).
The net degree of compatibility in $k_BT$ units
($\chi=n[\varepsilon_{+-}-(\varepsilon_{--}+\varepsilon_{++})/2]$, where
$n=6$ is the number of nearest neighbors) in our incompressible model is
simply $\chi=-3\varepsilon_{++}$, because for simplicity we only consider
short--range attraction among positively charged molecules (in a compressible
model all the interactions have to be included). It should be noted here that
$\chi$ describes the effective interaction among the co--assembled
macromolecules and not among single monomers. In the systems with only van
der Waals interactions $\chi$ is proportional to $1/T$, however, systems with
hydrophobic and hydrogen bonding interactions, such as in co-assembled PA's,
the pair--interaction may have a complex $T$ dependence. Excluded volume and
electrostatic interactions are also considered. The Hamiltonian of the system
reads as follows,
\begin{equation}
\frac{H}{k_BT}=\sum\limits_{i>j}^{N}\frac{Q_iQ_j}{\epsilon_{\rm o} r_{ij}}+
\sum\limits_{\{i,j\}}^{N}\frac{\varepsilon_{++}(Q_i+Q_+)(Q_j+Q_+)}{(2Q_+)^2},
\end{equation}
where $Q_{i}$ is a charge of the $i-$th unit, $\epsilon_{\rm o}$ is the
average dielectric constant of the media in units $e^2/4\pi\sigma$,
$r_{ij}=|\vec{r}_j-\vec{r}_i|/\sigma$ is a dimensionless distance between
$i-$th and $j-$th units, $\vec{r}_i$ determines the position of the $i-$th
unit in the space, $Q_+=1$ and the summation in the second term is only over
nearest neighbors. To exclude interaction with displaced image--cylinders,
periodical boundary conditions are applied only in the $z-$direction and,
thus, Lekner summation technique~\cite{GrzybowskiMolPhys2002} is used to
calculate the electrostatic energy of the single cylinder system
\begin{eqnarray}
\frac{E_{el}}{k_BT}=\frac{1}{2\epsilon_{\rm
o}}\sum\limits_{i,j=0}^{N}\sum\limits^{\infty}_{m=-\infty}
\frac{Q_iQ_j}{\sqrt{\rho_{ij}^2+(z_{ij}+mL_z)^2}}
\end{eqnarray}
where 
$\rho_{ij}=\left[{x_{ij}^2+y_{ij}^2}\right]^{1/2}$ and for $m=0$ the terms
with $i=j$ are omitted. We report standard canonical Monte Carlo simulations
following the Metropolis scheme for various values of
$\varepsilon_{++}\in[-4.5,-0.5]$, $\epsilon_{\rm o}=10$,
$R_{c}/\sigma\in[1,8]$ and $L_{c}/\sigma=100$. Simple moves in the phase
space are performed by exchange of two randomly chosen particles. Each system
is equilibrated during $10^5$ MC steps per particle and another $10^5$ MC
steps are used to perform measurements. The equilibration process is
accompanied by a gradual decrease of temperature (temperature annealing) from
$T_{max}=10$ to $T_{min}=1$. To analyze equilibrium properties we calculate
the heat capacity $C_V$ and the static structure factor
\begin{equation}
S(q)=\frac{1}{N_{+}}\left<\left|
\sum\limits_{i,j=0}^{N_{+}}e^{\imath\vec{q}\cdot(\vec{s}_j-\vec{s}_i)}\right|^2\right>,
\end{equation}
where $\vec{s}_{j}$ is a two--dimensional cartesian surface vector and $S(q)$
is averaged among different directions. We focus our attention on the size of
the segregated domains and on the susceptibility, a degree of correlations in
the system.

\begin{figure}
\includegraphics[width=8.4cm]{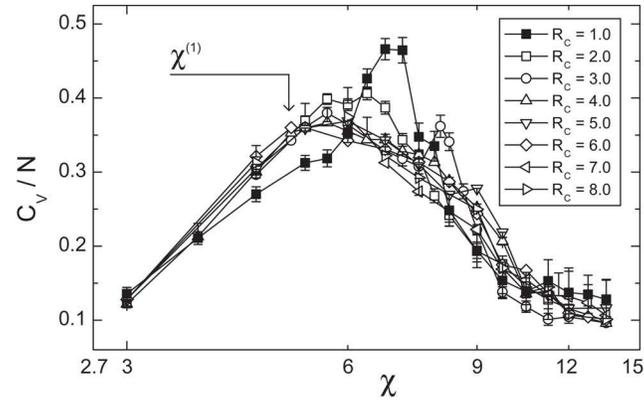}
\caption{\label{fig-2}Dimensionless, normalized by~$N,$ heat capacity as a
function of the pair attraction energy~$\chi$ for different $R_c$.}
\end{figure}
\begin{figure*}
\includegraphics[width=17cm]{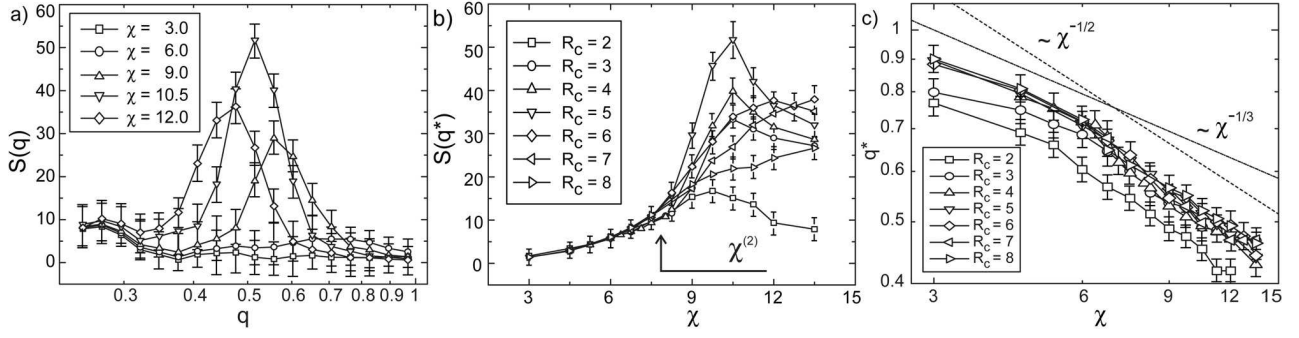}
\caption{\label{fig-3}(a) Static structure factor for $R_c=5$ for
different~$\chi$, (b) magnitude of the peak of the structure factor,
$S(q^*)$, as a function of~$\chi$ for different $R_c$ values, (c) double
logarithmic plot of the peak position $q^*$ versus~$\chi$ for different
$R_c$.}
\end{figure*}

The transition from the isotropic to the striped phase begins with the
appearance of segregated domains~(Fig.~\ref{fig-1}~(a)) and proceeds through
an intermediate locally correlated state governed by
fluctuations~(Fig.~\ref{fig-1}~(b)). In the planar (2D-monolayer) case, in
the limit of zero charge, a macroscopic phase segregation in the mean field
approximation is expected at a critical point~$\chi^{cp}\simeq 2$, as
predicted by a simple coarse--grained free energy functional of the neutral
system,~\cite{CahnJCP1958}
\begin{eqnarray}
\frac{\Delta F\left(\phi\right)}{k_BT}=
\int\frac{f(\phi)+\kappa|\nabla\phi|^2}{\sigma^2}d^2r, \label{free}
\end{eqnarray}
where $\phi=\phi\left(\mathbf{r}\right)$ is the local charge concentration,
$f\left(\phi\right)=\phi\ln\phi+\left(1-\phi\right)\ln\left(1-\phi\right)-
\chi\phi^2$, $\kappa\approx\chi a^2/2$~\cite{McCoy1976}. In the charged case
there is no critical point. Instead, a broad peak in the heat
capacity~(Fig.~\ref{fig-2}) at a characteristic value denoted by~$\chi^{(1)}$
is observed. The charges restricts the possibility of macroscopic segregation
and induces instead favorable fluctuations of wave length $\lambda$. If the
charge fluctuations are small, a density variation till second order in the
free energy in terms of the Fourier components of the density $\phi_q$ gives
$\Delta F/k_BT\sim\sum_q\phi_q\phi_{-q}/(2S_0(q))$, where $S_0^{-1}(q) = 4 -
2\chi + \chi q^2 + 8\pi/q\epsilon_{\rm o}$. The competition between
electrostatic,~$8\pi/q\epsilon_{\rm o}$, and gradient,~$\chi q^2$, energies
results in the formation of most favorable fluctuations of finite wave
length~$\lambda=2\pi/q^*$ and the appearance of a peak in ~$S_0(q)$
at~$q^*=(4\pi/\epsilon_{\rm o}\chi)^{1/3}$~as shown in the computed $S(q)$ in
Fig.~\ref{fig-3}~(a). This scaling regime is possible for~$\chi<\chi^{(1)}$
(fluctuations are expected to modify this mean field scaling)
(Fig.~\ref{fig-3}~(c)). On the other hand, in the high~$\chi$ limit
(low~$T$), strongly segregated charged domains of size $\Lambda$ with a well
defined line tension~$\gamma\sim\chi$~\cite{Chaikin} develop. The competition
of the interfacial energy among the segregated domains,~$\gamma \Lambda$, and
the electrostatic penalty associated with creating a charge
domain,~$(\Lambda^2)^2/(\epsilon_{\rm o} \Lambda),$ gives another scaling
limit~$\Lambda\sim(\epsilon_{\rm o}\chi)^{1/2}$~\cite{SolisJCP2005}. The
second scaling regime is found to exist for~$\chi>\chi^{(2)},$ where
$\chi^{(2)}$ is defined below.

The mean--field analysis of Eq.~\ref{free} erroneously predicts a continuous
transition to a periodic structure signaled by a mean field susceptibility
$S_0(q^*)=\infty$ at $\chi_{cp}=2+(3/2)[4\pi\sqrt{\chi_{cp}}/\epsilon_{\rm
o}]^{2/3}$.
It is well known that the transitions from isotropic to periodic structures
cannot be continuous in 3D~\cite{BrazovskiiJETF1975}. In 3D the
electrostatically driven microphase separation~\cite{DormidontovaMThS1994} is
often suppressed by various effects~\cite{YethirajJCP2003}. In 2D, since
there is no long range order~\cite{Peierls}, the possibility of a classical
thermodynamic transition to periodic structures is
questionable~\cite{FrenkelPRE2004}. In 1D, on the other hand, fluctuations
destroy the possibility of either macrophase or nano-phase transitions at non
zero~$T$~\cite{onedim}. We expect to recover the 2D limit only
when~$R_c\rightarrow\infty$ and the 1D case when~$R_c\rightarrow 0$.

\begin{figure*}
\includegraphics[width=17.0cm]{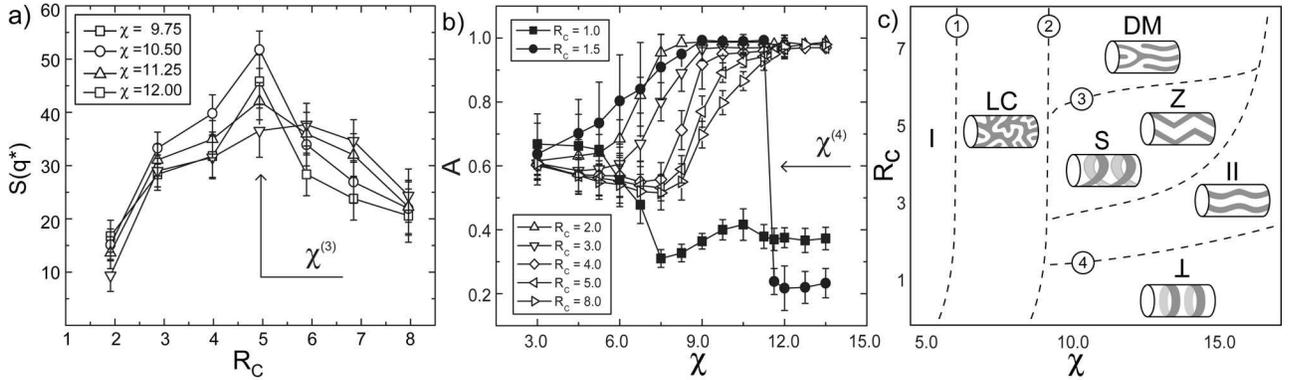}
\caption{\label{fig-4}a) Susceptibility $S(q^*)$ as a function of~$R_c$ near
extremum, b) asphericity order parameter as a function of~$\chi$ and~$R_c$,
c) schematic phase diagram $(\chi,R_c)$ implied by results of simulation,
where I -- isotropic phase, II -- parallel lamella phase, $\bot-$~ring phase,
LC -- locally correlated state, S -- helically twisted state, Z -- zigzag
state, DM -- defect mediated state. Lines $(1)$, $(2)$, $(3)$ and $(4)$
correspond to $\chi^{(1)}$, $\chi^{(2)}$, $\chi^{(3)}$ and $\chi^{(4)}$.}
\end{figure*}

Appearance of large domains on the surface of the cylinder of final radius
applies geometrical restrictions on the size and symmetry of the domains,
especially in the 1D--limit. These geometrical restrictions may influence
physical properties, such as the susceptibility. Figure~\ref{fig-3}~(b) shows
a non-monotonic dependence of~$S(q^*)$ with~$\chi$. We find that~$S(q^*)$ is
independent of~$R_c$ for~$\chi<\chi^{(2)}$, while for~$\chi>\chi^{(2)}$ it
shows a non monotonic dependence on~$R_c$. The point $\chi^{(2)}$ corresponds
to the cross--over transition from the state, where the segregated domains
are locally correlated, to the striped state, where the domain size scales
as~$\Lambda\sim(\epsilon_{\rm o}\chi)^{1/2}$ and the domains morphology
depends on~$R_c$. The melting of striped structures in the
limit~$R_c\rightarrow\infty$ may occur via the Kosterlitz--Thouless
mechanism~\cite{Nelson}. Besides the appearance of dislocations and other
defects~(Fig.~\ref{fig-1}~(c)), the striped state on the cylinder is richer
due to the possibility of different symmetries. We find a broad peak
in~$S(q^*)$ versus~$R_c$~(Fig.~\ref{fig-4}~(a)) for each~$\chi>\chi^{(2)}$,
implicitly denoted by~$\chi^{(3)}$, that moves to larger~$R_c$ values with
increasing~$\chi$. For intermediate~$\chi>\chi^{(2)}$ the structures are
strongly dependent on~$R_c$. For very wide cylinders $(q^*R_c>2.5)$ we find a
defect mediated striped state. With decreasing~$R_c$ the striped state turns
into the spiral/zigzag state~$(1<q^*R_c<2.5)$. With further decreasing~$R_c$
the spiral/zigzag phase, first, turns into a parallel lamella and then into
the ring phase~$(q^*R_c<1)$. With increasing the energy~$\chi$, the spiral
state turns into the zigzag state and then into the parallel striped phase.
Our simulation shows that for large~$\chi$ the parallel stripe phase
dominates unless the~$q^*R_c\sim 1$.

We calculate an order parameter with the purpose of characterizing the
geometry of the domains and to find their preferable morphology. We begin by
defining independent clusters on the surface of the cylinder. For each
cluster~$\mathcal{C}$ we calculate the inertia matrix with components
\begin{equation}
T_{\alpha\beta}\{\mathcal{C}\}=
\frac{1}{N_{\mathcal{C}}^2}\sum\limits_{i>j}^{N_{\mathcal{C}}}
\left(r^{\alpha}_i-r^{\beta}_i\right) \left(r^{\alpha}_j-r^{\beta}_j\right),
\end{equation}
where~$N_{\mathcal{C}}$ is the number of ions in the~$\mathcal{C}-$cluster
and~$r^{\alpha}_i$ is the~$\alpha-$th cartesian component of the position
vector of the~$i-$th ion. To characterize the anisotropy of the cluster we
use the "asphericity parameter" introduced by Rudnick and
Gaspar~\cite{RudnickGaspari}
\begin{equation}
A=\left\langle\frac{\left(Tr[T]\right)^2-3M[T]}
{\left(Tr[T]\right)^2}\right\rangle_{\mathcal{C}},
\end{equation}where $(R_1^2, R_2^2, R_3^2)$ are the eigenvalues of the matrix
$T_{\alpha\beta}$ which defining the three principal axes of the cluster,
$Tr[T]=R_{1}^2+R_{2}^2+R_{3}^2$ is the trace,
$M[T]=R_{1}^2R_{2}^2+R_{1}^2R_{3}^2+R_{2}^2R_{3}^2$ is the sum of three
minors and $A$ is averaged among all clusters. For spherically symmetric
objects $(R_{1}=R_{2}=R_{3})$ $A$ is equal to zero, for thin toroidal objects
$(R_{1}=R_{2}, R_{3}\ll R_{1})$ it is equal to~$A\simeq 1/4$ (in our case,
ring phase) and for long thin stripes~$(R_{2}\ll R_{1}, R_{3}\ll R_{1})$
$A\simeq 1$ (parallel lamella phase). Figure~\ref{fig-4}~(b) shows~$A$ as a
function of~$\chi$. For small values of~$\chi,$ when system is isotropic, the
asphericity parameter takes value~$A\simeq 0.6$ independent on~$R_c$. For
wide cylinders~$(q^*R_c>1)$ $A$ grows with~$\chi$ and approaches unity at
large~$\chi$ values. For relatively thin cylinders,~$q^*R_c\simeq 1.5,$ it
approaches unity faster than for cylinders with~$q^*R_c\simeq 2.5$, due to
the fact that $A$ does not distinguish stripes in parallel, spiral or zigzag
states. Meanwhile, for the stripes disturbed by the defects~$(q^*R_c\geq
2.5)$ the asphericity parameter takes smaller values than unity. In the case
of thin cylinders~$(q^*R_c<1)$ the asphericity parameter~$A$ indicates
formation of ring stripes. The transition from the parallel lamellar phase to
the ring phase is found at $\chi$ denoted by~$\chi^{(4)}$.

Our results are summarized in Fig.~\ref{fig-4}~(c), where we show a schematic
$(\chi,R_c)$ phase diagram. We predict various nano--phases obtained as a
result of the interplay of short-- and long--range forces among
stoichiometric charged components confined on the surface of a cylindrical
fiber. Our study suggests that besides modifying the size and the chemistry
of the molecules to induce different domain symmetries, one can co-assemble
basic and acidic molecules with different total charge per molecule, e.g.
$+2$ and $-1$, which may induce other domain symmetries along the surface of
the cylinder~\cite{SolisJCP2005}. Various different symmetries and
interesting phenomena appear as a result of the interplay between the
symmetry of the surface domains and the geometry of the aggregates. The
constraint of tiling the surface of aggregates with finite size domains of
different symmetries is a topic of interest and is relevant to studies of
viral symmetry and other self--assembled supra--molecular structures.

We acknowledge valuable discussions with S. I. Stupp, F. J. Solis
and A. Kudlay. This work is supported by NSF grant numbers
DMR-0414446 and DMR-0076097, and by NIH grant number GM62109.

\newpage


\end{document}